\documentclass[
twocolumn ,amssymb, amsmath,nobibnotes, aps, prl]{revtex4}

\expandafter\ifx\csname package@font\endcsname\relax\else
 \expandafter\expandafter
 \expandafter\usepackage
 \expandafter\expandafter
 \expandafter{\csname package@font\endcsname}%
\fi

\def\be{\begin{equation}}
\def\ee{\end{equation}}
\def\bea{\begin{eqnarray}}
\def\eea{\end{eqnarray}}

\begin{document}

\title{Topological Field Configurations in the Presence of Isospin Chemical Potential}
\author{M. Loewe}
\email{mloewe@fis.puc.cl} \affiliation{Facultad de F\'\i sica,
Pontificia Universidad Cat\'olica de Chile,\\ Casilla 306,
Santiago 22, Chile.}
\author{S. Mendizabal}
\email{smendizabal@fis.puc.cl} \affiliation{Facultad de F\'\i
sica, Pontificia Universidad Cat\'olica de Chile,\\ Casilla 306,
Santiago 22, Chile.}
\author{J. C. Rojas}
\email{jcristobalrojas@hotmail.com}
\affiliation{Departamento de F\'\i sica, Universidad Cat\'olica del Norte, Casilla 1280,\\
Antofagasta,  Chile.}

\date{\today}

\begin{abstract}
We analyze the stability of different topological solutions in
Quantum Field Theory when an isospin chemical potential $\mu $ is
included. We work in the limit when temperature vanishes. We find
that static vortex solutions in $2+1D$ do exist. However, the 't
Hooft-Polyakov monopole in $3+1D$ is no longer stable, as soon as
the chemical potential acquires a finite value. In the case of the
Skyrmion, this topological solution still exists for finite $\mu$,
up to a certain critical value.
\end{abstract}

\maketitle

\vskip1cm

The possibility of a new mechanism leading to spontaneous symmetry
breaking induced by the Bose-Einstein condensation, due to
chemical potentials, has recently been explored in the frame of
the electro-weak model \cite{sannino}. There is a crucial
difference with the standard symmetry breaking mechanism since now
the number of Nambu-Goldstone bosons that appear is lesser than
the number required by the Goldstone theorem \cite{miransky}.

In this letter, we explore the effect of the isospin chemical
potential on some topological structures that appear in Quantum
Field Theory (QFT). First, we will analyze the possibility  of
having vortex-type solutions  in a $U(1)$ invariant theory in
$2+1$ dimensions. Later, we address the question of the existence
of 't Hooft-Polyakov monopole solutions in $3+1$ dimensions,
finding that this kind of solutions are not longer topological
stable, since the homotopy group becomes trivial for finite
chemical potential values. Finally, we concentrate on Skyrmions,
showing that, contrary to the monopole solutions, the chemical
potential does not prevent their existence.

\section{Vortex in 2+1D}

Let us consider a complex field $\phi $ in $2+1$ dimensions

\be \phi=\phi_1+ i \phi_2, \; \; \phi = \left( \begin{array}{c}\phi_1 \\
 \phi_2\end{array} \right),
 \ee

\noindent with a $U(1)$-invariant lagrangian
 \be {\cal L} =
\partial_{\mu} \phi^{\ast} \partial_{\mu} \phi -\frac{\lambda}{2}
\left( \phi^{\ast} \phi - F^2 \right)^2. \label{2}
\ee

Following 't Hooft \cite{thooft}, the vacuum is defined through
the manifold:

\be
\mid \!\! \phi \!\! \mid = F.
\ee

Given a certain possible solution

\be
\mid \!\! \vec{x} \!\! \mid \rightarrow \infty :
\vec{\phi} = F \frac{\vec{x}}{\mid \!\! x \!\! \mid},
\phi= F e^{i \phi},
\ee

\noindent the energy is given by

\be E= \int d^2x \left( \vec{\partial} \phi^{\ast}\vec{\partial}
\phi + V(\phi,\phi^{\ast}) \right), \ee

\noindent which diverges, since

\be \mid \!\! \vec{x}\!\! \mid \rightarrow \infty : (\partial_i
\phi_j)^2 \rightarrow \frac{F^2}{\mid \!\! x \!\! \mid^2}. \ee

\noindent Then

\be
\int d^2 x \vec{\partial} \phi^{\ast}\vec{\partial} \phi
= 2 \pi \int d\mid \!\! x \!\! \mid \frac{F^2}{\mid \!\! x \!\! \mid^2}
\sim \mbox{div log}.
\ee

The previous divergence can be avoided through the introduction of
 a covariant derivative.

\be
\partial_{\mu} \rightarrow D_{\mu}\phi =
\left( \partial_{\mu} -i e A_{\mu}
\right) \phi.
\ee

The usual asymptotic finite energy solution, a vortex, is given by

\bea \phi &=& \Omega F, \; \; \; \Omega = e^{i \varphi},
\nonumber \\
\vec{A} &=& - \frac{1}{ie} \Omega \vec{\partial} \Omega^{-1},
\nonumber \\
A_i &=& - \frac{1}{e} \epsilon_{ij} \frac{x_j}{r^2}. \eea

\noindent To get a better understanding, we can analyze this
solution in polar coordinates

\bea
\vec{A}=\left(A_r,A_{\varphi} \right)
\rightarrow \left(0, \frac{1}{er} \right).
\eea

\noindent Using

\be \vec{A}=-\frac{1}{ie} e^{i\varphi} \vec{\partial}
e^{-i\varphi} = \frac{1}{e} \vec{\varphi},
\ee

\noindent the magnetic flux $\vec{B}=F_{12}$ inside a circle of
area $s$ is

\bea \Phi &=& \int_s \vec{B} \cdot d\vec{\sigma},
\nonumber \\
&=& \int_{c=\partial s} \vec{A} \cdot d\vec{l} = \int_{c=\partial
s} \frac{1}{e} \vec{\nabla} \varphi \cdot d\vec{l},
\nonumber \\
&=& \frac{1}{e}
\int_{c=\partial s} d\varphi = \frac{2 \pi}{e}.
\eea

\noindent This means that we have a quantized magnetic flux, in
units of   $g_m= 2 \pi/e$.

\noindent The energy of this solution is

\bea
E&=& \int d^2 x  \left[ D_i \phi^{\dagger} D_i \phi
+ \frac{1}{2}F^2_{12} +
\frac{\lambda}{2}(\phi^{\dagger}\phi-F^2)^2 \right]
\nonumber \\
&=& \int d^2 x  \left[ (\partial_i \phi)^2
+e^2 \vec{A}^2 \phi^2
+ \frac{1}{2}F^2_{12} +
\frac{\lambda}{2}(\phi^{\dagger}\phi-F^2)^2 \right].
\nonumber
\eea

\noindent Choosing the following parameter values, since we know
that
 $[\lambda]\sim [e^2]$ y $[eF]\sim m$ from dimensional analysis:

\be \lambda=e^2, \; \; \; m= m_{\phi}=m_A=\sqrt{2}e F,
\label{couplings}
 \ee

we have

\bea
E&=&\int d^2x  \left[
\left(
\partial_i \phi \pm e \epsilon_{ij}A_j \phi
\right)^2
\right.
\nonumber \\
&& \left. + \frac{1}{2} \left( F_{12} \pm \sqrt{\lambda}
(\phi^2-F^2) \right)^2 \pm e F^2 F_{12} \right],
\nonumber \\
&\geq& eF^2 \mid \int d^2x F_{12} \mid = e F^2 \frac{2
\pi}{e}=\frac{\pi m^2}{e^2}. \label{energy}
\eea

Let us introduce our chemical potential in the model. Now, the
lagrangian density becomes modified in the usual way \cite{Weldon}
according to

\be
{\cal L} =  {\cal D}_{\mu} \phi^{\ast} {\cal D}_{\mu} \phi
-\frac{\lambda}{2} (\phi^{\ast} \phi)^2 ,
\ee

\noindent with

\be {\cal D}_i = D_i, \;   {\cal D}_0 \phi = \partial_0\phi - i
\mu \phi, \;  {\cal D}_0 \phi^{\ast} = \partial_0\phi^{\ast} + i
\mu \phi^{\ast}, \ee

\noindent choosing  $A_0=0$, which is by far a non trivial gauge
election, the lagrangian density is

\bea
{\cal L}&=& D_{\mu}\phi^{\ast} D^{\mu}\phi
-\frac{\lambda}{2} \left[
(\phi^{\ast}\phi)^2 - 2 \frac{\mu^2}{\lambda}\phi^{\ast}\phi
+ \left( \frac{\mu^2}{\lambda} \right)^2
\right]
\nonumber\\
&&+ \frac{\mu^4}{2 \lambda}.
\eea

Neglecting the last constant term, and  noticing that the chemical
potential plays the same role as the term $F$ in (\ref{2}), we
have that, for the same election of (\ref{couplings}), there is a
nontrivial minimum for the energy:

\be
 E \geq 2\pi \left(
\frac{\mu^2}{\lambda} \right).
\ee

\noindent It is interesting to remark that the chemical potential
can be considered as a possible source of symmetry breaking.

\section{'t Hooft-Polyakov Monopole Solution (3+1D)}

Let us consider in what follows the $SO(3)$ invariant lagrangian

\be {\cal L}= -\frac{1}{4} F^a_{\mu \nu}F^{a \mu \nu} +\frac{1}{2}
D_{\mu}\phi^a D^{\mu}\phi^a -V(\phi), \ee

\noindent where we introduced the definitions

\bea
F^a_{\mu \nu} &=&
\partial_{\mu}A^a_{\nu}-\partial_{\nu}A_{\mu}
+e \epsilon^{a b c}A^b_{\mu}A^c_{\nu},
\nonumber \\
D_{\mu}\phi^a &=& \partial_{\mu} \phi^a +
e \epsilon^{a b c}A^b_{\mu} \phi^c,
\nonumber \\
V(\phi)&=& \frac{\lambda}{4}\left( \phi^a \phi^a - F^2 \right)^2.
\eea

\noindent The energy is given by

\be
E= \int d^3x \left(
\frac{1}{2}({\vec D}\phi^a)^2 +
\frac{\lambda}{4}(\phi^a \phi^a - F^2)
+\frac{1}{2} ({\vec B}^a)^2
\right).
\ee

\noindent Asymptotically, the monopole-type solution points
radially. In fact, the following ansatz satisfies this condition.

\bea \phi^a &=& \frac{r^a}{er^2} H(eFr),
\nonumber \\
A^a_0 &=& 0,
\nonumber \\
A^a_i &=& -\epsilon^{a i j}\frac{r_j}{er^2}[1-K(eFr)]. \eea

\noindent Notice that this solution is only invariant under the
diagonal subgroup of $SO(3)_R \otimes SO(3)_G$. In the previous
expressions, the functions $H$ and $K$ obey the boundary
conditions

\bea K(\xi)&\rightarrow& 1, \; \; H(\xi)\rightarrow 0, \mbox{ when
} \xi \rightarrow 0,
\nonumber \\
K(\xi)&\rightarrow& 0, \; \; \frac{H(\xi)}{\xi} \rightarrow 1,
\mbox{ when } \xi \rightarrow \infty, \eea

\noindent with $\xi=eFr$.

\noindent With this solution, the energy is given by

\bea
E &=& \frac{4 \pi F}{e} \int^{\infty}_0
\left[
\xi^2 \left( \frac{dK}{d\xi}
\right)^2 + K^2 H^2
+\frac{1}{2} \left(\xi \frac{dH}{d\xi}-H
\right)^2
\right.
\nonumber \\
&& \left.
+ \frac{1}{2} \left(
K^2-1
\right)^2
+ \frac{\lambda}{4 e^2} \left(
H^2 - \xi^2
\right)^2
\right].
\eea

According to the usual recipe, the lagrangian density when isospin
chemical potential is introduced, becomes modified as

\bea
{\cal L} &=& -\frac{1}{4} F^a_{\mu \nu}F^{a \mu \nu}
+\frac{1}{2} D_{\mu}\phi^a D^{\mu}\phi^a
-V(\phi)
\nonumber\\
&&
+ \frac{\mu^2}{2} (\phi^2_1+\phi^2_2).
\eea

\noindent The last two terms contribute to the energy density
according to

\bea \label{break}\Delta {\cal E}&=& V(\phi)-\frac{\mu^2}{2}
(\phi^2_1+\phi^2_2), \nonumber
\\
&=& \frac{\lambda}{2}\left( \vec{\phi}^2-\zeta^2 \right)^2 +
\frac{\mu^2}{2}\phi_3^2, \eea

\noindent with

\be
\zeta^2= \frac{F^2}{2}+\frac{\mu^2}{4\lambda}.
\ee

\noindent Note that the last term in (\ref{break}) breaks
explicitly the diagonal $SO(3)$ symmetry. In order to have a
finite energy solution, we need

\be \phi_3(r\rightarrow\infty)=0,
(\phi_1^2+\phi_1^2)(r\rightarrow\infty)=\zeta^2,
\ee

\noindent i.e. the vacuum manifold reduces to $S_1$. Since the
homotopy group $\pi_2(S_1)=0$, we conclude that there is no room
for the existence of 't Hooft-Polyakov monopoles in the presence
of a finite isospin chemical potential, which is equivalent to a
charge asymmetric state of matter.

\noindent It is possible, nevertheless, to have an accidental
monopole solution when the original potential becomes

\be V(\phi)=\frac{\lambda}{2}\left( \vec{\phi}^2
\right)^2-\frac{m^2}{2} \phi_3^2, \ee

\noindent when $\mu=m$. In this particular case the rotational
field symmetry is restored.

\section{Skyrmions}

Skyrme's original idea \cite{skyrme} was to consider the baryons
as solitons in the non-linear sigma model. Later, this idea was
fairly  well supported by the predictions of static properties of
baryons \cite{nappi}, being the baryonic number explained as a
topological effect \cite{witten}. Extensions of this idea to the
Hybrid models, the Cheshire Cat scenarios at finite temperature
were also considered \cite{falomir}.

Due to the phenomenological success of the Skyrme model, it is
natural to analyze the influence of the isospin chemical potential
on the Skyrmion solution.

\noindent The lagrangian of the non-linear sigma model with the
stabilizing skyrme term is given by

\bea {\cal L} &=& \frac{F_{\pi}^2}{16} Tr\left[
\partial_{\mu}U \partial^{\mu}U^{\dagger}
\right] \nonumber \\ &&+ \frac{1}{32e^2}Tr \left[
(\partial_{\mu}U)U^{\dagger}, (\partial_{\nu}U)U^{\dagger}
\right]^2. \eea

As the chemical potential plays the role of a temporal gauge
field, it is natural the generalization of the covariant
derivative:

\be
\partial_{\nu} U \rightarrow  D_{\nu}U = \partial_{\nu}U -i \frac{\mu}{2} [\sigma^3,U] g_{\nu
0}.
\ee

\noindent Defining $L_{\nu} \equiv (\partial_{\nu}U) U^{\dagger}$,
 the lagrangian density becomes

\bea {\cal L}_{\mu} &=&
 \frac{F_{\pi}^2}{16} Tr \left\{
 \partial_{\nu}U \partial^{\nu}U^{\dagger} + \frac{\mu^2}{2}
\left[ \sigma_0 - U \sigma_3 U^{\dagger} \sigma_3 \right] \right\}
\nonumber \\
 &+& \frac{1}{32e^2} Tr\left\{ \left[L_{\rho},L_{\nu} \right]^2 -
\frac{\mu^2}{2} \left( [\varrho,L_{\nu}][\varrho,L^{\nu}] \right)
\right\}, \eea

\noindent where $\varrho = \sigma_3-U\sigma^3 U^{\dagger}$.

For static solutions, we obtain

\bea {\cal L}_{\mu} &=& {\cal L}_{\mu=0} + \frac{F_{\pi}^2
\mu^2}{32} Tr \left[ \sigma_0 - U \sigma_3 U^{\dagger} \sigma_3
\right]
\nonumber\\
&& + \frac{\mu^2}{64 e^2} Tr  \left[ \varrho,L_{\nu} \right]^2.
\eea

\noindent We parameterize the field $U$ matrix as

\be U = \exp \left( \xi \frac{\vec{\sigma}}{2i}\cdot \hat{n}
\right)=\cos\frac{\xi}{2} -i(\vec{\sigma} \cdot
\hat{n})\sin\frac{\xi}{2}. \ee

\noindent The ansatz by Skyrme has an ``Hedgehog" shape and the
following boundary conditions have to be satisfied

\bea &&\xi(\vec{r})=\xi(r), \; \; \hat{n}=\hat{r},
\nonumber \\
&&\xi(0)=2 \pi, \; \; \xi(\infty)=0. \eea

\noindent where $\xi(r)$ can be solved variationally. It is
possible to find an approximate analytical solution \cite{atiyah},
in terms of dimensionless parameters

\be \label{atiyah} \xi= 2 \pi \left[ 1-
\frac{r}{\sqrt{\lambda^2+r^2}}\right]. \ee

\noindent The optimal value for $\lambda$, from minimizing the
mass of the Skyrmion is $\lambda=1.453$.

The chemical potential dependent mass is given by

\be M_{\mu}=M_{\mu=0}- \frac{\mu^2}{4 e^3 F_{\pi}} I_2 -
\frac{\mu^2}{32 e^3 F_{\pi}} I_4,\ee

\noindent where the dimensionless integrals $I_2$ and $I_4$ are

\bea I_2 &=& \int d^3r Tr \left[ \openone
-U\sigma_3U^{\dagger}\sigma_3\right],
\nonumber \\
I_4 &=&  \int d^3r Tr \left[ \varrho,L_{\nu} \right]^2.\eea

Notice that the corrections coming from the chemical potential
terms have a negative sign. This means that eventually, for some
critical value of $\mu_c$, the Skyrmion mass should vanish. This
picture is attractive, since it suggests that nucleons disappear
for critical values of density.

As an approximation, we will estimate these integrals using the
ansatz (\ref{atiyah}), and the value of $M_{\mu=0}=36.5
\frac{F_{\pi}}{e}$ given by \cite{witten}, obtaining

\be M_{\mu}= 36.5 \frac{F_{\pi}}{e}-33.72 \frac{\mu^2}{e^3
F_{\pi}}. \ee

\noindent This gives a critical value

\be \mu_c=1.04 \; e  F_{\pi} \approx 731 MeV .\ee

This value is rather high. However, we have presented only a first
rough discussion of this issue. Later we will present a more
detailed analysis of Skyrmion stability. Our results implies that
nucleons, according to the Skyrmion description will be stable in
heavy ion collisions and other scenarios where we expect to
achieve isospin chemical
potential values of the order of the pion mass.\\

\begin{acknowledgments}
M.L. and S.M. acknowledge support from a FONDECYT grant under
number 1010976. S.M. acknowledges also support from DIPUC.
\end{acknowledgments}

\end{document}